\documentstyle[12pt]{article}
\newcommand{\bm}[1]{\mbox{\boldmath $#1$}}
\newcommand{\rd}{{\rm d}}
\newcommand{\be}{\begin{equation}}
\newcommand{\ee}{\end{equation}}
\newcommand{\ba}{\begin{eqnarray}}
\newcommand{\ea}{\end{eqnarray}}
\newcommand{\bb}[1]{\bibitem{#1}}
\begin{document}
\begin{titlepage}
\setcounter{page}{1}
\title{Hollow cosmic string: the general-relativistic hollow
 cylinder}
\author{G\'erard Cl\'ement\thanks{E-mail:
 GECL@CCR.JUSSIEU.FR} \\
\small Laboratoire de Gravitation et Cosmologie Relativistes
 \\
\small Universit\'e Pierre et Marie Curie, CNRS/URA769 \\
\small Tour 22-12, Bo\^{\i}te 142 --
 4, place Jussieu, 75252 Paris cedex 05, France
\and
Ilhem Zouzou\\
\small D\'epartement de Physique Th\'eorique --
 Universit\'e de Constantine\\
\small route d'Ain-el-Bey, DZ-25000 Constantine, Alg\'erie}
\bigskip
\date{\small June 1, 1994}
\maketitle
\begin{abstract}
We determine the different possible space-time metrics
 inside an infinite rotating hollow cylinder with given
 energy density and longitudinal and azimuthal stresses, the
 metric outside the cylinder being chosen of the spinning
 cosmic string type. The solutions we obtain for various
 domains of values of the cylinder parameters include a
space-time with topologically Euclidean spatial sections, a
 black-hole solution, a quasi-regular solution, and various
 wormhole solutions. A solution which is regular only if the
 longitudinal dimension is compactified might approximately
 describe spontaneous compactification of the cylinder to a
 torus.
\end{abstract}
\end{titlepage}
\section{Introduction}
The space-time metric at large distance from a thin,
 straight, infinitely long cosmic string is well known
\cite{1,9} to be flat but non-Minkowskian, with an angular
 deficit proportional to the linear energy density of the
 string. On the other hand, the gravitational field in the
 region of space where the energy-momentum tensor is non-negligible
 depends on the precise mechanism responsible for
 the existence of the string. A limiting case which may be
 of some interest is
 that of a hollow cosmic string, the energy-momentum tensor
 being concentrated not on a line but on the surface of an
 infinitely long cylinder. The gravitational field generated
 by such a cylinder with given energy density and
 longitudinal and azimuthal stresses, and rotating about its
 axis, was in principle determined some time ago by Frehland
 \cite{2}. One of the results of Frehland's analysis is
that,
 unexpectedly, a rotating hollow cylinder can exist in
 general relativity only if its energy-momentum is
 traceless. Actually, as we shall show below, this
 intriguing result is not general, but follows from the
 hidden assumption that the apparent singularities of the
 space-time metric outside and inside the cylinder coincide.
 Once this unnecessary constraint is removed, Frehland's
 `traceless' condition no longer holds.

This remark led us into reanalyzing the problem of the
 rotating hollow cylinder, along the lines of the analysis
 carried out in \cite{3} for a rotating closed string in
 three-dimensional space-time. The space-time metric outside
 and inside the cylinder is a stationary, cylindrically
 symmetric solution of the vacuum Einstein equations. We
 first give, in the second section of this paper, a complete
 classification of these solutions. We expect that, from a
 large distance, the hollow cylinder should look like a thin
 straight cosmic string, so it is natural to choose the
 metric outside the cylinder to be the well-known conical
 metric. Making this choice, we apply in the third section
 the junction method \cite{4} to derive the parameters of
 the interior metric from the knowledge of the cylinder's
 energy-momentum tensor, which we assume to have the
 anisotropic perfect-fluid form. The generic vacuum
 stationary cylindrically-symmetric metric being singular on
 a cylindrical surface $S_-$ coaxial with the matter
 cylinder, the interior metric we have obtained can be
 regular in two cases. Either 1) the singular surface $S_-$
 is in the interior region, but the singularity is only a
 coordinate singularity. This possibility, studied in the
 fourth section, includes, along with a solution with global
 spatial sections which are topologically ${\bf R}^3$ ($S_-$
 then reduces to the axis of symmetry), two other families
 of non-topologically Euclidean solutions: black cosmic
 string solutions, and a family of quasi-regular solutions,
 including a limiting case which we interpret as
 corresponding
 to a closed hollow cosmic string. Or 2) the singular
 cylinder $S_-$ lies outside the matter cylinder, so that
 the singularity is only virtual. As discussed in the fifth
 section, the corresponding solutions can be divided in two
 classes, according to the fate of test particles falling
 inside the cylinder, which may either be eventually
 reflected back to the ouside space, or accelerated towards
 the cylinder at interior spatial infinity. We close with a
 brief discussion.
\setcounter{equation}{0}
\section{Stationary cylindrically symmetric vacuum space-times}
A stationary cylindrically symmetric metric has three
 commuting Killing vectors, one of which is time-like. Such
 a metric can be parametrized in adapted coordinates by
\be
\rd s^2 = \lambda_{ab} (r) \, \rd x^a \, \rd x^b - \rd r^2
\ee
($a,b=0,1,2$), with $x^0=t$, $x^1= \varphi$ ($0 \leq \varphi
 < 2 \pi$), $x^2=z$. With this parametrization, the Einstein
 equations read \cite{6}
\ba
S^r_{\,\,\,r} & \equiv & \frac{1}{8} \, \left[ {\rm Tr}(
 \chi ^2) - ({\rm Tr} \chi)^2 \right] = \kappa \,
 T^r_{\,\,\,r} \, , \nonumber \\
S^a_{\,\,\,b} & \equiv & \frac{1}{2} \, \left[ \chi
 '^a_{\,\,\,\,b} + \frac{1}{2} \, ({\rm Tr} \chi) \, \chi
 ^a_{\,\,\,b} \right. \nonumber \\
& & \mbox{} - \left. \delta ^a_b \, \left( {\rm Tr} \chi ' +
 \frac{1}{4} \, \left[ {\rm Tr}( \chi ^2) + ({\rm Tr}
 \chi)^2 \right] \right) \right] = \kappa \, T^a_{\,\,\,b}
\ea
(${}'= \rd / \rd r$), where $\chi$ is the $3 \times 3$
 matrix
\be
\chi = \lambda ^{-1} \, \lambda ' \, .
\ee

Surprisingly, although particular solutions of the vacuum
 equations (2.2) ($T^\mu _{\,\,\, \nu}=0$) have been
 previously discussed in the literature \cite{6,5,10}, no
 exhaustive classification of these solutions is, to our
 knowledge, available. In vacuum, the equations (2.2)
 simplify to
\ba
& & {\rm Tr}(\chi^2) = ({\rm Tr} \chi)^2 \, , \\
& & \chi ' + \frac{1}{2} \, ({\rm Tr} \chi) \, \chi = 0 \,.
\ea
Putting
\be
f \equiv \frac{1}{2} \, {\rm Tr} \chi \, ,
\ee
and tracing eq.\  (2.5), we obtain the differential equation
\be
f' + f^2 = 0 \,
\ee
which is solved either by
\be
f = \frac{1}{r-r_0} \, ,
\ee
or by
\be
f = 0 \, .
\ee

In the first, generic case, the full eq.\  (2.5) is solved
 by
\be
\chi = \frac{2}{r-r_0} \, A
\ee
where $A$ is a constant $3 \times 3$ matrix, which according
 to eqs.\  (2.6) and (2.4) is constrained by
\be
{\rm Tr} A = {\rm Tr} A^2 = 1 \, .
\ee
The integration of eq.\  (2.3) then leads to
\be
\lambda = C \, \exp \, [2 \, A \, \ln |r-r_0|] \, ,
\ee
where $C$ is a new constant matrix, of signature ($+ - -$).
The symmetry of the matrix $\lambda$ implies the symmetry
 relations
\be
C^T = C \, , \,\,\,\, (C A)^T = C A \, .
\ee
The constraints (2.11) are equivalent to the relations
\be
p_0 + p_1 + p_2 = 1 \, , \,\,\,\, {p_0}^2 + {p_1}^2 +
 {p_2}^2 = 1 \, ,
\ee
between the three eigenvalues of the matrix $A$. If these
 three eigenvalues are real and different, the matrix $A$
 may be brought into the diagonal form
\be
A = \sum_i p_i \, n^{(i)} \otimes \overline{n}^{(i)} \, ,
\ee
where the $n^{(i)}$ are the linearly independent
 eigenvectors of $A$, and the $\overline{n}^{(i)}$ are the
 dual basis one-forms such that
\be
\langle \overline{n}^{(i)} , n^{(j)} \rangle \equiv
 \overline{n}^{(i)}_a \, n^{(j)a} = \delta _{ij} \, .
\ee
The symmetry properties (2.13) imply that the matrix $C$ is
 diagonal in the basis $n^{(i)} \otimes n^{(j)}$, so that
 eq.\  (2.12) reads
\be
\lambda = \sum_i \, k_i \, |r - r_0|^{2p_i} \,
 \overline{n}^{(i)} \otimes \overline{n}^{(i)} \, ,
\ee
leading to the Kasner-like \cite{7,16} metric
\be
\rd s^2 = \sum_i \, k_i \, |r - r_0|^{2p_i} \, (\rd y^i)^2 -
 \rd r^2 \, ,
\ee
where\footnote{Note that, owing to the periodicity condition
 on $x^1=\varphi$, (2.19) is not in general a globally
 defined coordinate transformation. As a consequence, the
 metric (2.18) is generically stationary, not static.}
\be
\rd y^i = \overline{n}^{(i)}_a \rd x^a \, .
\ee
If an eigenvalue $p$ is degenerate ($p=0$ or $p=2/3$), then
 $(A-p)^3 = \linebreak (1-3p)(A-p)^2$, so that (2.12) leads
 to
\ba
\lambda & = & C \, \left\{ \frac{(A-p)^2}{(1-3p)^2} \, |r-
r_0|^{2-4p} + \left[ 1 - \frac{(A-p)^2}{(1-3p)^2} \right] \,
 |r-r_0|^{2p} \right. \nonumber \\
& & \mbox{} + \left. \left[A - p - \frac{(A-p)^2}{1-3p}
 \right] \, \ln |r-r_0| \cdot |r-r_0|^{2p} \right\} \, ,
\ea
which is not of the Kasner form if $(A-p)^2 \neq (1-3p)(A-
p)$;
the case $p=0$ is more fully discussed in sect.\  4. If two
 eigenvalues of $A$ are complex conjugate, with real part
 $\rho$ and imaginary part $\sigma$, the Lagrange
 interpolation formula
\be
f(A) = \sum_i \, f(p_i) \, \prod_{j \neq i} \, \frac{A-
p_j}{p_i-p_j}
\ee
leads to
\ba
\lambda & = & M \, |r-r_0|^{2-4\rho} + [N \, \cos (2 \,
 \sigma \, \ln |r-r_0|) \nonumber \\
& & \mbox{} + P \, \sin (2 \, \sigma \, \ln |r-r_0|)] \, |r-
r_0|^{2\rho} \, ,
\ea
where the matrices $M, N, P$ are quadratic functions of $A$.
 The preceding cases are also discussed in \cite{6}.

In the second case $f=0$, eq.\  (2.5) is solved by
\be
\chi = A \, ,
\ee
where the constant matrix $A$ is now constrained by
\be
{\rm Tr}A = {\rm Tr}A^2 = 0 \, ,
\ee
which implies the matrix relation
\be
A^3 = (\det A) \, \bf{1}
\ee
(conversely eq.\  (2.25) implies eq.\  (2.24), unless $A$ is
 proportional to the unit matrix $\bf{1}$). Equation (2.3)
 then leads to
\be
\lambda = C \, {\rm e}^{Ar} \, ,
\ee
where the matrix $C$ has the same signature and symmetry
 properties as before. Non-diagonal solutions of (2.25) may
 be classified \cite{11} according to the rank $r(A)$ of the
 matrix $A$:

(a)  $\bm{r(A)} = {\bf 3}$. Then $A$ has the three
 eigenvalues $a$, $ja$, $j^2a$ (where $j={\rm e}^{2i \pi
 /3}$), leading to
\ba
\lambda & = & \frac{1}{3} \, C \, [(1 + \overline{A} +
 \overline{A}^2) \, {\rm e}^{ar} + (2 - \overline{A} -
 \overline{A}^2) \, {\rm e}^{-ar/2} \, \cos
 (\frac{\sqrt{3}}{2} \, ar) \nonumber \\
& & \mbox{} + \sqrt{3} \, (\overline{A}-\overline{A}^2) \,
 {\rm e}^{-ar/2} \, \sin (\frac{\sqrt{3}}{2} \, ar)] \, ,
\ea
where $\overline{A} = a^{-1}A$; this space-time is of Petrov
 type $I$.

(b)  $\bm{r(A)} = {\bf 2}$, or $A^3=0$, $A^2 \neq 0$. An
 example is
\be
A =
\left(
\begin{array}{rrr}
0 & 0 & 0 \\
0 & 0 & -1 \\
1 & 0 & 0
\end{array}
\right)
 \, , \,\,\, C =
\left(
\begin{array}{rrr}
0 & 1 & 0 \\
1 & 0 & 0 \\
0 & 0 & -1
\end{array}
\right)
 \, ,
\ee
leading to the metric
\be
\rd s^2 = - \frac{1}{2} \, (r \, \rd t + 2 \, \rd z)^2 + 2
 \, \rd t \, \rd \varphi + \rd z^2 -\rd r^2 \, .
\ee
The Petrov type is $N$.

(c)  $\bm{r(A)} = {\bf 1}$, or $A^2 = 0$. Choosing
\be
A =
\left(
\begin{array}{rrr}
0 & 0 & 0 \\
1 & 0 & 0 \\
0 & 0 & 0
\end{array}
\right)
 \, , \,\,\, C =
\left(
\begin{array}{rrr}
0 & 1 & 0 \\
1 & 0 & 0 \\
0 & 0 & -1
\end{array}
\right)
 \, ,
\ee
we obtain the metric \cite{3}
\be
\rd s^2 = r \, \rd t^2 + 2 \, \rd t \, \rd \varphi - \rd z^2
 -\rd r^2 \, .
\ee
In this case the curvature tensor vanishes, so that there
 must exist a local coordinate transformation which maps the
 metric into the Minkowski form
\be
\rd s^2 = \rd U \, \rd V - \rd Y^2 - \rd Z^2 \, .
\ee
For the choice (2.30), this transformation is
\ba
& & t = U \, , \,\,\,\, \varphi = \frac{1}{2} \, [V +
 \frac{1}{6} \, U^3 - U \, Y] \, , \nonumber \\
& & r = Y - \frac{1}{4} \, U^2 \, , \,\,\,\, z = Z \, .
\ea
However, because of the periodicity condition on the null
 coordinate $V$, the space-time of metric (2.31) is not
 globally equivalent to Minkowski space-time.

(d)  $\bm{r(A)} = {\bf 0}$, i.\  e.\  $A = 0$. The metric
 reduces to the ``cylindrical Minkowski metric''
\be
\rd s^2 = \rd t^2 - R^2 \, \rd \varphi^2 - \rd z^2 - \rd r^2
 \, ;
\ee
this is again locally, but not globally, equivalent to
 Minkowski space-time.
\setcounter{equation}{0}
\section{The junction method for the hollow cylinder
 problem}
We now return to our hollow cosmic string problem. Let us
 for convenience translate  the radial coordinate $r$ (which
 is defined by the parametrization (2.1) only up to an
 additive constant) so that the surface $r=0$ corresponds to
 the cylinder surface. As in Frehland's analysis of the
 hollow cylinder problem \cite{2}, we use the junction
 method \cite{4} to determine the space-time metric, i.\
 e.\  the $3 \times 3$ matrix $\lambda (r)$, from the
 knowledge of the energy-momentum tensor $T^\mu_{\,\,\,
 \nu}=\kappa^{-1} {\cal T}^\mu_{\,\,\, \nu} \, \delta (r)$
 concentrated on the surface of the matter cylinder. The
 metric field $\lambda (r)$ may be written
\be
\lambda (r) = \lambda_+ (r) \, \theta (r) + \lambda_- (r) \,
 \theta (-r) \, ,
\ee
where the exterior and interior metrics $\lambda_+$ and
 $\lambda_-$ solving the vacuum Eistein equations may always
 be chosen so as to smoothly join along the cylinder $r=0$:
\be
\lambda_- (0) = \lambda_+ (0) \, .
\ee
Because eq.\  (2.4) is satisfied everywhere, the energy-momentum
component $T^r_{\,\,\,r}$ vanishes on the cylinder
 from the first eq.\  (2.2). The remaining Einstein
 equations (2.2) then relate the discontinuity of the first
 derivative of the metric field to the surface energy-
momentum ${\cal T}$:
\be
\chi_+ (0) - \chi_- (0) = 2 \, {\cal T} - ({\rm Tr} {\cal
 T}) \, {\bf 1} \, .
\ee

We choose the exterior metric to be the flat ``spinning
 cosmic string'' metric , generalizing the three-dimensional
 ``spinning point particle'' metric of \cite{12},
\be
\rd s^2_+ = (\rd t - \omega \, \rd \varphi)^2 - \alpha^2 \,
 (r-r_+)^2 \, \rd \varphi^2 - \rd z^2 - \rd r^2 \, ,
\ee
so that a distant observer will see the cylinder as a cosmic
 string of mass per unit length $M=2 \pi (1- \alpha)/
 \kappa$ and spin per unit length $J=2 \pi \omega / \kappa$.
 This metric is of the form (2.17), with $p_1=1$,
 $p_0=p_2=0$, and the basis vectors and one-forms
\ba
& & n^{(0)}_+ =
\left(
\begin{array}{c}
1\\
0\\
0
\end{array}
\right)
 \, , \,\,\, n^{(1)}_+ =
\left(
\begin{array}{c}
\omega\\
1\\
0
\end{array}
\right)
 \, , \,\,\, n^{(2)}_+ =
\left(
\begin{array}{c}
0\\
0\\
1
\end{array}
\right)
 \, , \nonumber \\
& & \overline{n}^{(0)}_+ = \bm{(}1,- \omega ,0\bm{)} \, ,
 \,\,
\overline{n}^{(1)}_+ = \bm{(}0,1,0\bm{)} \, , \,\,
\overline{n}^{(2)}_+ = \bm{(}0,0,1\bm{)} \, .
\ea
Of course the metric (3.4) is regular in the exterior region
 only if the conical singularity is virtual, $r_+<0$. As for
 the interior metric, we only assume that it has the generic
 form
\be
\lambda _- = C_- \, \exp \, [2 \, A_- \, \ln |r-r_-|] \, ,
\ee
with the matrix $A_-$ constrained by (2.11); the various
 `exceptional' solutions (2.26) shall be recovered from
 (3.6) in the limit $r_- \rightarrow \infty$. Note that our
 assumptions are, in a certain sense, dual to those of
 Frehland \cite{2}, who assumed the interior metric to be
 Minkowskian, while the exterior metric was the general
 Kasner form (2.18).

Following Frehland, we assume the surface energy-momentum of
 the cylinder to be that of a uniformly rotating anisotropic
 perfect fluid:
\be
{\cal T} =
\left(
\begin{array}{cc}
(\mu - \tau_\varphi) \, u \otimes \overline{u} +
 \tau_\varphi \, {\bf 1} & 0\\
0 & \tau_z
\end{array}
\right) \, ,
\ee
with eigenvalues $\mu$ (surface energy density),
 $\tau_\varphi$ and $\tau_z$ (azimuthal and longitudinal
 stresses); here {\bf 1} is the $2 \otimes 2$ unit matrix.
 The local 2-velocity $u$ is normalized by $\langle
 \overline{u} , u \rangle = 1$, where $\overline{u} = u^T
 \lambda (0)$. We shall use the parametrization relative to
 the exterior basis
\be
u = (1- \beta^2)^{-1/2} \, (n^{(0)}_+ + \Omega \, n^{(1)}_+)
 \, , \,\, \overline{u} = (1 - \beta^2)^{-1/2} \,
 (\overline{n}^{(0)}_+ - \frac{\beta^2}{\Omega} \,
 \overline{n}^{(1)}_+) \, ,
\ee
where $\Omega$ is the angular velocity of the rotating
 cylinder, and $\beta=- \alpha r_+| \Omega |$ is its linear
 velocity.

With these assumptions, the Einstein equations (3.3) read
\ba
\lefteqn{\frac{1}{r_+} \, n^{(1)}_+ \otimes
 \overline{n}^{(1)}_+ - \frac{1}{r_-} \, A_- =} \nonumber \\
& & \left(
\begin{array}{cc}
(\tau_\varphi - \mu) \, u \otimes \overline{u} +
 \frac{\displaystyle (\mu - \tau_\varphi +
 \tau_z)}{\displaystyle 2} \, {\bf 1} & 0 \\
0 & \frac{\displaystyle (\mu + \tau_\varphi -
 \tau_z)}{\displaystyle 2}
\end{array}
\right) \, ,
\ea
leading to the form of the unknown matrix $A_-$ in the basis
 $n^{(i)}_+ \otimes \overline{n}^{(j)}_+$:
\be
A_- = -r_-
\left(
\begin{array}{ccc}
\frac{\tau_\varphi-\mu}{1-\beta^2} + \frac{(\mu -
 \tau_\varphi + \tau_z)}{2} &
\frac{\beta^2 \, (\mu - \tau_\varphi)}{\Omega \, (1-
\beta^2)} &
0 \\
\frac{(\tau_\varphi-\mu) \, \Omega}{1-\beta^2} &
\frac{\beta^2 \, (\mu - \tau_\varphi)}{1-\beta^2}+
 \frac{(\mu - \tau_\varphi + \tau_z)}{2} - \frac{1}{r_+} &
0 \\
0 &
0 &
\frac{(\mu + \tau_\varphi - \tau_z)}{2}
\end{array}
\right) \, .
\ee
This matrix must obey the two constraints (2.11), which lead
 to the two relations
\ba
& & \frac{1}{r_+} - \frac{1}{r_-} = \frac{1}{2} \, (\mu +
 \tau_\varphi + \tau_z) \, , \nonumber \\
& & \beta^2 = \frac{\tau_\varphi - r_+ \, (\mu \tau_\varphi
 - \frac{1}{4} \, (\mu + \tau_\varphi - \tau_z)^2)}{\mu -
 r_+ \, (\mu \tau_\varphi - \frac{1}{4} \, (\mu +
 \tau_\varphi - \tau_z)^2)} \, .
\ea
These equations may be solved to give the metric parameters
 $r_+$ and $r_-$ in terms of the cylinder parameters $\mu$,
 $\tau_\varphi$, $\tau_z$ and $\beta$:
\ba
& & r_+ = \frac{\tau_\varphi - \mu \, \beta^2}{(1-\beta^2)
 \, (\mu \tau_\varphi - \frac{1}{4} \, (\mu + \tau_\varphi -
 \tau_z)^2)} \, , \nonumber \\
& & r_- = \frac{4 \, (\tau_\varphi - \mu \,
 \beta^2)}{\beta^2 \, (3 \, \mu^2 + (\tau_\varphi -
 \tau_z)^2) - ((\mu - \tau_z)^2 + 3 \, \tau_\varphi^2)} \, .
\ea
Frehland's traceless condition \cite{2} ${\rm Tr} {\cal T} =
 \mu + \tau_\varphi + \tau_z = 0$ results from (3.11) if
 $r_+$ and $r_-$ happen to be equal. However, let us
 emphasize that this can only be a coincidence, as there is
 a priori no relation whatsoever between the locations of
 the virtual singularity $r_+$ of the exterior metric and of
 the (real or virtual) apparent singularity $r_-$ of the
 interior metric.

Solving the constraints (2.14) to determine the eigenvalues
 $p_0$ and $p_1$ of the matrix $A_-$ from the eigenvalue
 $p_2$ (obvious from eq.\  (3.10)), and taking into account
 the first relation (3.11), we obtain
\ba
& & p_{\stackrel{0}{1}} = \frac{1}{2} \, \frac{r_-}{r_+} \,
 (1 - r_+ \, \tau_z \mp \varepsilon \, \sqrt{\Delta}) \, ,
 \nonumber \\
& & p_2 = - \frac{1}{2} \, r_- \, (\mu + \tau_\varphi -
 \tau_z) \, ,
\ea
where
\be
\Delta = (1 - r_+ \, \tau_z) \, [1 - 2 \, r_+ \, (\mu +
 \tau_\varphi) + r_+ \, \tau_z] \, ,
\ee
and $\varepsilon$ is a sign which shall be determined below.
 The corresponding arbitrarily normalized eigenvectors and
 associated one-forms are
\ba
& & n^{(\stackrel{0}{1})}_- = \gamma \, n^{(0)}_+ +
 \frac{1}{2} \, (\eta \mp \varepsilon \, \sqrt{\Delta}) \,
 n^{(1)}_+ \, , \,\,\, n^{(2)}_- = n^{(2)}_+ \, , \nonumber
 \\
& & \overline{n}^{(\stackrel{0}{1})}_- = \pm \delta^{-1} \,
 [\frac{1}{2} \, (\eta \pm \varepsilon \, \sqrt{\Delta}) \,
 \overline{n}^{(0)}_+ - \gamma \, \overline{n}^{(1)}_+] \, ,
 \,\,\, \overline{n}^{(2)}_- = \overline{n}^{(2)}_+ \, ,
\ea
where we have put
\be
\gamma = \frac{r_+ \, (\tau_\varphi - \mu)}{\Omega} \,
 \frac{\beta^2}{1 - \beta^2} \, , \, \eta = 1 + r_+ \,
 (\tau_\varphi - \mu) \, \frac{1+\beta^2}{1-\beta^2} \, , \,
 \delta = \frac{1-\beta^2}{1+\beta^2} \, \gamma \,
 \varepsilon \, \sqrt{\Delta} \, .
\ee
To obtain the interior metric in the form (2.17), there
 remains to enforce the continuity conditions (3.2), which
 may be written as $\lambda_- (0) n^{(i)}_- = k_i |r_-
|^{2p_i} \overline{n}^{(i)}_- = \lambda_+ (0) n^{(i)}_-$,
 leading to
\ba
& & k_{\stackrel{0}{1}} = \pm \frac{\varepsilon}{2} \,
 \alpha^2 \, r^2_+ \, \sqrt{\Delta} \, (\eta \mp \varepsilon
 \, \sqrt{\Delta}) \, |r_-|^{-2p_{\stackrel{0}{1}}} \, ,
 \nonumber \\
& & k_2 = - |r_-|^{-2p_2} \, .
\ea
It can be checked that the product $k_0k_1$ is negative, as
it should. The anholonomic coordinate $y_0$ in (2.18) is
timelike if $k_0$ is positive, which corresponds to the
choice
\be
\varepsilon = {\rm sign} \, \eta
\ee

The metric we have obtained is a solution to the hollow
cylinder problem only if it is everywhere regular (except
for the discontinuity (3.3) on the cylinder surface). Now
the interior metric has an apparent singularity on the
cylinder $r=r_-$. If $r_-<0$, a necessary condition for
regularity is that the metric coefficients be analytic
functions of $r$. This is possible if the exponents $p_i$
are integer, i.\  e.\  from eq.\  (2.14) if one of the $p_i$
is equal to 1 and the two others vanish. For $p_1=1$, the
 spatial sections have the Euclidean topology, while the two
other cases $p_0=1$ and $p_2=1$ have non-Euclidean spatial
topologies; we take the point of view that an observer
outside the string should not rule out the possibility of
exotic interior topologies. The other possibility is $r_-
>0$, in which case the interior metric is regular for all
values of the $p_i$. In this case (in which we also include
 the limit $r_- \rightarrow \infty$, leading to exceptional
interior solutions), the global topology is that of a
Lorentzian wormhole, with the two axes at infinity $r
\rightarrow \pm \infty$. We study these various
possibilities in the following two sections.
\setcounter{equation}{0}
\section{Solutions with an apparent singularity inside the
cylinder}

(a)  $\bm{p}_{\bf 1} = {\bf 1}$. Then $p_2=0$ which, from
the last equation (3.13), is possible only if the cylinder
parameters are related by
\be
\tau_z = \mu + \tau_\varphi \, .
\ee
In this case (as well as in case (b) $p_0=1$, which also
implies $p_2=0$), the matrix $A_-$ of eq.\  (3.10) is really
two-dimensional\footnote{It then follows from eq.\  (2.11)
that $A^2_-=A_-$, so that the interior metric can be put in
the Kasner form even though the eigenvalue $0$ is
degenerate.} so that, taking into account the continuity of
the metric tensor, the interior metric reduces to
\be
\rd s^2_- = \lambda_{-(2)ab}(r) \, \rd x^a \, \rd x^b - \rd
z^2 - \rd r^2 \, .
\ee
The global space-time is then the product of the $z$ axis by
the three-dimensional space-time generated by a rotating
closed string of energy density $\mu$ and stress
$\tau_\varphi$ \cite{3}.

The first equation (3.13) with $p_0=0$, $p_1=1$ leads to the
conditions
\be
1 - r_+ \, \tau_z = \varepsilon \, \sqrt{\Delta} =
\frac{r_+}{r_-} \, .
\ee
As $r_+$ and $r_-$ are both negative, it follows that
$\varepsilon=+1$. The interior metric may be written in the
form (2.18),
\be
\rd s^2_- = \rd t'^2 - (r-r_-)^2 \, \rd \varphi ' - \rd z^2
- \rd r^2 \, ,
\ee
with
\ba
& & \rd t' = \sqrt{k_0} \, \delta^{-1} \, [\frac{1}{2} \,
(\sqrt{\Delta}+\eta) \, \rd t - (\frac{1}{2} \,
(\sqrt{\Delta}+\eta) \, \omega +
\frac{\delta}{\sqrt{\Delta}}) \, \rd \varphi] \, , \nonumber
\\
& & \rd \varphi' = \sqrt{-k_1} \, \delta^{-1} \,
[\frac{1}{2} \, (\sqrt{\Delta}-\eta) \, \rd t - (\frac{1}{2}
\, (\sqrt{\Delta}-\eta) \, \omega -
\frac{\delta}{\sqrt{\Delta}}) \, \rd \varphi] \, .
\ea
However this coordinate transformation is consistent with
the periodicity condition on $\varphi$ only if the time $t'$
does not depend on $\varphi$ and if the angle $\varphi'$ has
the same period $2\pi$ as $\varphi$, i.\  e.
\be
\frac{1}{2} \, (\eta + \sqrt{\Delta}) \, \omega +
\frac{\delta}{\sqrt{\Delta}} = 0  \, , \,\,\, \sqrt{-k_1} \,
\delta^{-1} \, [\frac{1}{2} \, (\eta - \sqrt{\Delta}) \,
\omega + \frac{\delta}{\sqrt{\Delta}}] = 1
\ee
(if these conditions are not satisfied, then the interior
metric is a spinning cosmic string metric, with a conical
singularity at $r=r_-$). The transformation (4.5) then
reduces to a time dilation coupled with a uniform frame
rotation of angular velocity $-\omega / r^2_-$.

Because the interior metric (4.4) is Minkowskian, the proper
radius of the cylinder $r=0$ is unambiguously identified as
$|r_-|$, which is different from the two possible exterior
definitions $|r_+|$ and $\alpha |r_+|$. Taking the
independent parameters of the cylinder to be $|r_-|$, $\mu$
and $\tau_\varphi$, $\tau_z$ is given by (4.1), while from
eqs.\  (3.11) the linear rotation velocity of the cylinder
is
\be
\beta = \frac{\tau_\varphi \, (1 + r_- \, \tau_\varphi)}{\mu
\, (1 + r_- \, \mu)}
\ee
(positive from eq.\  (4.9) below). If such a fine tuning
between the cylinder parameters is not achieved, then the
centrifugal energy of the rotating cylinder does not balance
its azimuthal stress, and the cylinder either contracts or
expands. The external metric parameter $r_+$ is, from the
first eq.\  (3.11),
\be
r_+ = \frac{r_-}{1 + r_- \, (\mu + \tau_\varphi)} \, .
\ee
The remaining exterior metric parameters $\alpha$ and
$\omega$ are then obtained from the regularity conditions
(4.6),
\ba
& & \alpha^2 = (1 + r_- \, \mu) \, (1 + r_- \, \tau_\varphi)
\, (1 + r_- \, (\mu + \tau_\varphi)) \, , \nonumber \\
& & \omega^2 = \frac{r_-^4 \, \mu \, \tau_\varphi}{1 + r_-
\, (\mu + \tau_\varphi)}
\ea
(to obtain this last relation, we have also used the
definition $\beta = -\alpha r_+ |\Omega|$). Combining eqs.\
(4.8) and (4.9), we obtain the relation
\be
\alpha^2 \, r_+^2 - \omega^2 = r_-^2 \, .
\ee
In (4.10) we recognize the continuity condition (3.2) for
the metric component $g_{\varphi \varphi}$ (which is
unaffected by a uniform frame rotation). A consequence is
that $g_{\varphi \varphi}$ stays negative outside as well as
inside the cylinder $r=0$ so that, contrary to the case of
the spinning cosmic string space-time \cite{12}, the global
space-time generated by the rotating cylinder does not
contain closed timelike curves. This is a very satisfactory
feature of this solution.

We still have to enforce the inequalities $r_+<0$, $r_-<0$,
$\omega^2 \geq 0$ (from which follows $\alpha^2>0$) and
$\beta<1$. These inequalities lead to two possible domains
of parameter values. The first possibility
\be
0 \leq \tau_\varphi < \mu  \, , \,\,\, \mu + \tau_\varphi <
\frac{1}{|r_-|}
\ee
also implies
\be
0 \leq \tau_\varphi < \beta^2 \, \mu \, ,
\ee
and leads to the total cylinder mass per unit length
\be
M = \frac{2 \, \pi}{\kappa} \, (1 - \alpha) > 2 \, \pi \,
|r_-| \, \frac{\mu}{\kappa} \, .
\ee
The second possibility $\mu<\tau_\varphi \leq 0$, which also
implies $\beta \mu < \tau_\varphi < \beta^2 \mu$, must be
excluded as it leads to a negative net mass per unit length
$M < 2 \pi |r_-| \mu / \kappa <0$.

(b)  $\bm{p}_{\bf 0} = {\bf 1}$. In this case again, $p_2=0$
implies the relation (4.1), while eq.\ (4.3) is here
replaced by
\be
1 - r_+ \, \tau_z = -\varepsilon \, \sqrt{\Delta} =
\frac{r_+}{r_-} \, ,
\ee
from which follows $\varepsilon = -1$. This last condition
is equivalent to the inequality $k_{0-}>0$, which may be
written
\be
(1 - r_+ \, \mu) \, (1 - r_+ \, \tau_\varphi) <0 \, .
\ee
Taking into account eqs.\  (4.7) and (4.8), which hold again
in this case, the inequality (4.15) results in the condition
\be
0 \leq - \beta \, \mu < \tau_\varphi \, .
\ee

The conditions (4.9) have no equivalent here. The reason is
that, in this case, a suitable uniform frame rotation on the
interior metric determined in the previous section leads to
the metric \cite{3}
\be
\rd s^2_- = \nu^2 \, (r-r_-)^2 \, (\rd t - \omega _- \, \rd
\varphi')^2 - R^2 \, \rd \varphi'^2 - \rd z^2 - \rd r^2 \, .
\ee
The apparent singularity of this metric at $r=r_-$ is actually
a cylindrical horizon. The corresponding space-time
is flat, as may be shown by carrying out the local Rindler-like
\cite{14} transformation
\ba
\overline{t} & = & (r-r_-) \, \sinh \, [\nu \, (t - \omega_-
\, \varphi')] \, , \nonumber \\
\overline{r} & = & (r-r_-) \, \cosh \, [\nu \, (t - \omega_-
\, \varphi')] \, , \nonumber \\
\overline{\varphi} & = & \varphi' \, , \,\,\, \overline{z} =
z \, ,
\ea
which maps the domain $r_- \leq r \leq 0$ of the interior
space-time on one of the two domains $I$ ($0 \leq
\overline{r}^2-\overline{t}^2 \leq r^2_-$) of the Minkowski
cylinder
\be
\rd s^2_- = \rd \overline{t}^2 - \rd \overline{r}^2 - R^2 \,
\rd \overline{\varphi}^2 - \rd \overline{z}^2 \, .
\ee
The maximal extension of the metric (4.17) is of course
diffeomorphic to the entire Minkowski cylinder. However, the
maximally extended interior space-time is not globally
equivalent to the Minkowski cylinder, for two reasons. The
first is that the Rindler time $t-\omega_-\varphi'$ in
(4.17) has a helical structure \cite{12}; a consequence is
the occurence of closed timelike curves outside the cylinder
$\nu^2 \omega^2_- (r-r_-)^2=R^2$. The second reason is the
inequality $r\leq 0$, which results in the truncation
$\overline{r} \leq (r^2_- + \overline{t}^2)^{1/2}$ of the
Minkowski cylinder.

The metric (4.17) is invariant under the discrete symmetry
$r-r_- \rightarrow r_- - r$. The full space-time geometry is
invariant under this symmetry if the matter cylinder at
$r=0$ is echoed by a mirror image cylinder at $r=2r_-$. The
Penrose diagram of this space-time is shown in Fig.\  1. The
two mirror-symmetric cylinders separate space-time into
three regions, two symmetrical exterior regions with a
truncated spinning cosmic string geometry, and an interior
region with a truncated spinning Rindler cylinder geometry
 (the truncated maximal extension of the metric (4.17)). An
observer falling inside the stationary cylinder may send
lightlike signals to the outside until he crosses the
horizon $r=r_-$. Thereafter the observer continues falling
through flat space-time towards interior timelike infinity,
while the cylindrical walls now seem, in the Minkowskian
frame (4.19), to expand according to $\overline{r} = \pm
(r^2_- + \overline{t}^2)^{1/2}$. The attempts of the trapped
 observer to communicate with the outside are now doomed, as
his signals $\overline{r}-\overline{t} = {\rm constant}$
never catch up with the receding walls, and end up at
interior lightlike infinity ${\cal I}^+_{in}$. The future
Rindler region $II$ ($\overline{r}^2 - \overline{t}^2<0$) is
thus, for an external observer, a cylindrical black hole
with event horizon $r=r_-$ \cite{15}.

We should emphasize here that this black hole structure
arises if the parameters of the matter cylinder are
constrained by (4.1) and (4.16), so that the matter making
up the cylinder has negative energy density $\mu$ (even in
the static limit $\beta \rightarrow 0$) and positive
stresses $\tau_\varphi$ and $\tau_z$. The exotic character
of this matter shows up in the dynamics (the fate of a
falling observer), but not in the exterior space-time
 geometry, which is the spinning cosmic string geometry
(3.4) with parameters $\alpha$ and $\omega$ which may be
arbitrarily chosen (only $r_+$ is determined from the string
parameters by eq.\  (3.12)). We may thus choose these
parameters so as to ensure that the total (matter +
gravitational) mass per unit length $M=2\pi (1-
\alpha)/\kappa$ is positive ($\alpha<1$) and that there are
no timelike curves ($|\omega|<\alpha |r_+|$, which also
 ensures the absence of closed timelike curves inside the
matter cylinder, because of the continuity of $g_{\varphi
\varphi}$).

(c)  $\bm{p}_{\bf 2} = {\bf 1}$. From eqs.\  (3.13) and
(3.11), this occurs if the metric parameters $r_+$ and $r_-$
are related to the longitudinal tension $\tau_z$ by
\be
r_+ = \frac{1}{\tau_z} \, , \,\,\, r_- = -\frac{2}{\mu +
\tau_\varphi - \tau_z} \, ,
\ee
while the linear rotation velocity $\beta$ of the string is
\be
\beta = \left| \frac{\mu - \tau_\varphi - \tau_z}{\mu -
\tau_\varphi + \tau_z} \right| \, .
\ee
The inequalities $r_+<0$, $r_-<0$ ($r_-$ finite) and
$\beta<1$ are equivalent to the conditions
\be
\mu < \tau_\varphi \, , \,\,\, \tau_z < 0 \, , \,\,\, \tau_z
< \mu + \tau_\varphi
\ee
on the cylinder parameters.

On account of eqs.\  (4.20) and (4.21), the matrix $A_-$ of
eq.\  (3.10) reduces to
\be
A_- = \left(
\begin{array}{ccc}
a & \varepsilon' \, \frac{\displaystyle
\alpha}{\displaystyle \tau_z} \, a & 0 \\
- \varepsilon' \, \frac{\displaystyle \tau_z}{\displaystyle
\alpha} \, a & -a & 0 \\
0 & 0 & 1
\end{array}
\right) \, ,
\ee
where
\be
a = \frac{\tau_z^2 - (\mu - \tau_\varphi)^2}{2 \, \tau_z \,
(\mu + \tau_\varphi - \tau_z)} \, ,
\ee
and $\varepsilon' = {\rm sign}(\Omega [\tau_z^2 - (\mu -
\tau_\varphi)^2])$. According to eq.\  (2.20), the resulting
matrix $\lambda_-$ is
\be
\lambda_- = C_- \, [1-A_-^2+(A_--A_-^2) \, \ln \, (r-r_-
)^2+A_-^2 \, (r-r_-)^2] \, ,
\ee
with
\be
A_-^2 = \left(
\begin{array}{ccc}
0 & 0 & 0 \\
0 & 0 & 0 \\
0 & 0 & 1
\end{array}
\right) \, .
\ee
If $a \neq 0$, the matrix $A_--A_-^2$ is non zero so that
the metric corresponding to (4.25) is non-Kasner. Using the
continuity condition (3.2) to determine the matric $C_-$, we
obtain the interior metric
\ba
\rd s^2_- & = & (\psi + 1) \, \rd t^2 - 2 \, (\omega' \psi +
\omega) \, \rd \varphi \, \rd t + [\omega'^2 \psi + \omega'
\, (2 \, \omega - \omega')] \, \rd \varphi^2 \nonumber \\
& & \mbox{} -\frac{(r-r_-)^2}{r_-^2} \, \rd z^2 - \rd r^2 \,
,
\ea
with
\be
\psi = a \, \ln \, \frac{(r-r_-)^2}{r_-^2} \, , \,\,\,
\omega' = \omega + \varepsilon' \, \alpha \, r_+ \, ,
\ee

The only non-zero components of the curvature tensor
following from (4.27) are
\be
R^{\alpha 2}_{\,\,\,\,\,\,\beta 2} = -R^{\alpha
3}_{\,\,\,\,\,\,\beta 3} = -\frac{1}{(r-r_-)^2} \,
A^{\alpha}_{\,\,\, \beta}
\ee
($\alpha, \beta = 0,1$), of Petrov type $I$. Obviously the
curvature invariants vanish, so that the singularity of the
interior metric at $r=r_-$ is a non-scalar curvature
singularity \cite{17}. Because of this singularity, the
metric (4.27) is not a regular solution to the hollow
cylinder problem. However, as we shall now show by studying
the geodesic motion of a test particle in this metric, for a
certain range of the cylinder parameters the singularity
$r=r_-$ is `harmless', i.\  e.\  it is avoided by all
timelike and lightlike geodesics. The geodesic equation in a
stationary cylindrically symmetric metric
(2.1) may be integrated to
\be
\left( \frac{\rd r}{\rd \tau} \right)^2 - \Pi^T \,
\lambda^{-1} \, \Pi + \eta = 0 \, ,
\ee
where $\eta = +1$, $0$ or $-1$, $\tau$ is an affine
parameter, and the generalized momenta
\be
\Pi_a = \lambda_{ab} \, \frac{\rd x^a}{\rd \tau}
\ee
are constants of motion. In the case of the metric (4.27),
equation (4.30) reads
\be
\left( \frac{\rd r}{\rd \tau} \right)^2 + p^2 \, (\psi +1) +
2 \, p \, \Pi_0 + \Pi^2_2 \, \frac{r_-}{(r-r_-)^2} +\eta = 0
\, ,
\ee
with $p = -(\omega' \Pi_0 + \Pi_1) \varepsilon'/\alpha r_+$.
All the geodesics are deflected away from $r=r_-$ by the
potential barrier in $(r-r_-)^{-2}$, unless $\Pi_2=0$. If
$\Pi_2=0$, then for $a>0$ almost all the geodesics terminate
at the singularity $r=r_-$ ($\psi \rightarrow - \infty$).
However for $a<0$ all the geodesics are deflected away by
the logarithmic potential barrier $p^2 \psi$, unless $p=0$.
In this last case ( which excludes timelike geodesics $\eta
= 1$), only spacelike geodesics ($\eta = -1$) extend to
$r=r_-$. We conclude that test particles are always repelled
by the singularity $r=r_-$ if $a<0$, i.\  e.\
\be
\mu > \tau_\varphi + \tau_z
\ee
(note that the constraints (4.22) and (4.33) allow the
possibility of a positive energy density $\mu$). An
unpleasant corollary of the choice $a<0$ is that for $\psi$
large enough the interior geometry (4.27) admits closed
timelike lines. This pathology does not occur if the
parameters of the exterior metric are constrained by
$\omega' = 0$ ($\omega=-\varepsilon' \alpha r_+$), which
also excludes closed timelike lines outside the cylinder.
However the interior geometry still admits for $\omega' = 0$
closed lightlike lines, the closed null geodesics $\rd t =
\rd z = \rd r = 0$ ($\Pi_1 = \Pi_2 = 0$).

If now $a=0$, which from eq.\  (4.21) implies
\be
\tau_z = \mu - \tau_\varphi \, , \,\,\, \beta = 0 \, ,
\ee
the matrix $A_-$ assumes the diagonal form (4.26), and the
interior metric is simply
\be
\rd s^2_- = (\rd t - \omega \, \rd \varphi)^2 - \alpha^2 \,
r^2_+ \, \rd \varphi^2 - \frac{(r-r_-)^2}{r^2_-} \, \rd z^2
- \rd r^2 \, .
\ee
This metric has a conical singularity at $r=r_-$, unless the
variable $z/|r_-|$ is an angle, i.\  e.\  unless $z$ and
$z+2\pi |r_-|$ are identified. To be consistent, such an
identification should be carried over to the exterior
metric, so that the `longitudinal' dimension $z$ in (3.4) is
actually wrapped up around a circle. This implies that our
hollow cosmic string is not cylindrical but toroidal, with
small radius $\alpha |r_+| = -\alpha/\tau_z$ and large
radius $|r_-|=1/\tau_\varphi$. Of course, space-time around
such a torus can only be approximately described by the
cylindrical metric (3.4), a necessary condition for such an
approximation being $|r_-| \gg \alpha |r_+|$, i.\  e.\
\be
\alpha \, \tau_\varphi \ll - \tau_z \, .
\ee
The hollow cosmic string then connects two three-dimensional
spaces where the torus $r=0$ is embedded in two different
ways, the exterior space with approximate metric (3.4), and
the interior space (4.35) where the torus now has $|r_-|$ as
`small' radius and $\alpha |r_+|$ as `large' radius.

Weak-field self-gravitating circular cosmic strings have
been constructed in \cite{18}. Our various hollow cosmic
string solutions can presumably be approximately extended to
the case of circular cosmic strings by compactifying the
longitudinal dimension $z$. The solution $p_2=1$, $a=0$ is
unique in that it {\em spontaneously} compactifies. It is
weak-field if $\alpha \simeq 1$, in which case
$0<\tau_\varphi \ll -\mu$ from eqs.\  (4.34) and (4.36), so
that the energy density on the torus is negative. Contrary
to the result of \cite{18}, the total mass of this hollow
cosmic string
\be
2 \, \pi \, |r_-| \, M = (1-\alpha) \, \frac{4 \,
\pi^2}{\kappa} \, |r_-|
\ee
is different from the matter energy $4 \pi^2 |r_-| \alpha
|r_+| \mu / \kappa \simeq - \alpha (4 \pi^2 / \kappa) |r_-
|$. The corresponding gravitational radius $(1-\alpha)(\pi
/2) |r_-|$ is, in the weak-field case $\alpha \simeq 1$,
negligible (as it should) with respect to the radius $|r_-|$
of the matter distribution.
\setcounter{equation}{0}
\section{Solutions without an apparent singularity}

The interior metric (3.6) is regular for $r \leq 0$,
whatever the values of the Kasner exponents $p_i$, if $r_-
>0$. In this case the two regularity conditions $r_+<0$,
$r_->0$ constrain the possible domains of values of the
cylinder parameters. To simplify the notation, let us define

\be
x = \mu + \tau_\varphi \, , \,\,\, y = \mu - \tau_\varphi \,
, \,\,\, z = \tau_z \, .
\ee
We have, from eqs.\  (3.11) and (3.12),
\be
r_+ = \frac{2}{t} \, , \,\,\, r_- = \frac{2}{t - x - z} \, ,
\ee
with
\be
t = \frac{y^2 + z^2 - 2 \, xz}{h^{-1}y - x} \, , \,\,\, h =
\frac{1 - \beta^2}{1 + \beta^2} \, .
\ee
Then the conditions $r_+<0$, $r_->0$ read
\be
x + z \leq t \leq 0 \, .
\ee

The nature of the interior geometry depends on the value of
\be
p_2 = \frac{z - x}{t - x - z} \, .
\ee
For $-1/3 \leq p_2 \leq 0$ the other two exponents $p_0$ and
$p_1$ are both positive; for $0 \leq p_2 \leq 1$, one of the
other exponents is positive and the other is negative;
finally, for $p_2 < -1/3$ or $p_2 >1$, the exponents $p_0$
and $p_1$ are complex conjugate. A lengthy analysis
\cite{19} of these inequalities together with the regularity
conditions (5.4) leads to the domains of parameter values
shown in Table 1, where we have put
\ba
& & k = \frac{1-\beta}{1+\beta} \, , \,\,\, l =
\frac{\sqrt{3} \, (1-\beta^2) + 2 \, \beta}{\sqrt{3} \, (1-
\beta^2) - 2 \, \beta} \, , \nonumber \\
& & y_0 = \sqrt{z \, (2 \, x - z)} \, , \,\,\,
\overline{y}_{\pm} = k^{\pm 1} \, (2 \, x - z) \, , \,\,\,
\tilde{y}_{\pm} = k^{\pm 1} \, z \, , \nonumber \\
& & y_{\pm} = \frac{1}{2} \, [h^{-1} \, (x + z) \pm
\sqrt{(h^{-2}-1) \, (x+z)^2 -3 \, (x-z)^2}] \, .
\ea
The various domains c are related to the corresponding
domains b by the involution $(x,y,z) \rightarrow (x,y,2x-
z)$.
\renewcommand{\arraystretch}{1.5}
\begin{table}
a) $\bm{p}_{\bf 2} \leq {\bf -1/3}$ \\
\(
\left.
\begin{array}{ccl}
\hspace*{4.4 mm} 1) & lx \leq z \leq hk^{-1}x \leq 0 \, , &
\beta \leq 1/\sqrt{3} \\
 & \left.
\begin{array}{c}
z \leq hk^{-1}x \leq 0 \\
x \geq 0 \, , \,\,\, z \leq lx
\end{array}
\right\}
& \beta \geq 1/\sqrt{3}
\end{array}
\right\} \,\,\,
y_- \leq y \leq y_+ \\
\begin{array}{clc}
\hspace*{5 mm} 2) & hk^{-1}x \leq z \leq x \leq 0 \, , & y_-
\leq y \leq \overline{y}_- \,\,\, {\rm or} \,\,\, y_+ \leq y
\leq \overline{y}_+ \\
\hspace*{5 mm} 3) & 2x \leq z \leq hk^{-1}x \leq 0 \, , &
\overline{y}_- \leq y \leq \overline{y}_+
\end{array}
\) \\ \\
b) ${\bf -1/3} \leq \bm{p}_{\bf 2} \leq {\bf 0}$ \\
\(
\begin{array}{cll}
\hspace*{5 mm} 1) & 2x \leq z \leq hk^{-1}x \leq 0 \, , & -
y_0 \leq y \leq \overline{y}_- \\
\hspace*{5 mm} 2) & hk^{-1}x \leq z \leq x \leq 0 \, , &
\overline{y}_- \leq y \leq -y_0 \\
\hspace*{5 mm} 3) & 2x \leq z \leq x \leq 0 \, , &
\overline{y}_+ \leq y \leq y_0
\end{array}
\) \\ \\
c) ${\bf 0} \leq \bm{p}_{\bf 2} \leq {\bf 1}$ \\
\(
\begin{array}{cll}
\hspace*{5 mm} 1) & hkx \leq z \leq 0 \, , & -y_0 \leq y
\leq \tilde{y}_- \\
\hspace*{5 mm} 2) & x \leq z \leq hkx \leq 0 \, ,
\,\,\,\,\,\,\,\,\, & \tilde{y}_- \leq y \leq -y_0 \\
\hspace*{5 mm} 3) & x \leq z \leq 0 \, , & \tilde{y}_+ \leq
y \leq y_0
\end{array}
\) \\ \\
d) ${\bf 1} \leq \bm{p}_{\bf 2}$ \\
\(
\begin{array}{cll}
\hspace*{5 mm} 1) & x \leq z \leq hkx \leq 0 \, , & y_- \leq
y \leq \tilde{y}_- \\
\hspace*{5 mm} 2) & x \leq 0 \, , \,\,\, hkx \leq z \leq
l^{-1}x \,\,\,
\left\{
\begin{array}{l}
\beta \leq 2 - \sqrt{3} \, , \\
\beta \geq 2 - \sqrt{3} \, ,
\end{array}
\right.
& \!\!\!
\begin{array}{l}
\tilde{y}_- \leq y \leq y_- \\
y_- \leq y \leq y_+
\end{array}
\end{array} \\
\left.
\begin{array}{cll}
\hspace*{4.4 mm} 3) &
x \leq z \leq l^{-1}x \leq 0 \, , & \,\,\,\, \hspace*{12 mm}
\beta \leq 2 - \sqrt{3} \\
& x \leq z \leq hkx \leq 0 \, , & \,\,\,\, \hspace*{12 mm}
\beta \geq 2 - \sqrt{3}
\end{array}
\right\} \,\,
y_+ \leq y \leq \tilde{y}_+ \\
\left.
\begin{array}{cll}
\hspace*{4.4 mm} 4) &
l^{-1}x \leq z \leq 0 \, , & \,\,\,\, \hspace*{20 mm} \beta
\leq 2 - \sqrt{3} \\
& hkx \leq z \leq 0 \, , & \,\,\,\, \hspace*{20 mm} \beta
\geq 2 - \sqrt{3}
\end{array}
\right\} \,\,
\tilde{y}_- \leq y \leq \tilde{y}_+
\)
\\ \\
\caption{The various domains of the cylinder parameter values
leading to regular wormhole solutions}
\end{table}
\renewcommand{\arraystretch}{1}

Because the interior metric is regular for all values of
$r\leq 0$, the space-times of this section have the wormhole
topology, with two axes at infinity $r \rightarrow \pm
\infty$. To ascertain whether these wormholes are
traversable or not, we study the motion of a test particle
in the interior metric, given by the integrated geodesic
equation (4.30) with $\eta = +1$ or $0$.

When the Kasner exponents $p_i$ are all real (domains b and
c), this equation reads
\be
\left( \frac{\rd r}{\rd \tau} \right) ^2 - \sum_i k^{-1}_i
\, \langle \Pi , n^{(i)}_- \rangle ^2 \, |r-r_-|^{-2p_i} +
\eta = 0 \, .
\ee
For large negative $r$, the leading term in the effective
potential comes from the Kasner term with negative exponent
$p_i$ (for $-1/3 \leq p_2 \leq 1$, one and only one of the
$p_i$ is negative). The wormhole is traversable if this term
is timelike ($p_0 < 0$), and is generically not traversable
if this term is spacelike ($p_1 <0$ or $p_2 <0$, so that
$p_0 > 0$). The sign of $p_0$ depends both on the sign of
$p_2$ and on the sign $\varepsilon$ in (3.13), which we can
show \cite{19} to be negative in the cases b1, b2, c1,
 c2, and positive in the cases b3, c3. We conclude that in
the cases b and c3, $p_0>0$ and the test particle is
generically reflected by a potential barrier in $|r-r_-|^{-
2p_j}$ ($j = 1$ or $2$), except if the constant of the
motion $\langle \Pi , n^{(j)}_- \rangle$ happens to vanish;
in this last case massive particles ($\eta = +1$) are again
reflected to exterior spacelike infinity, while massless
particles ($\eta = 0$) can travel to interior spacelike
 infinity $r \rightarrow - \infty$ in the cases b1, b2 and
in the case c3 if $p_2 > 2/3$. In the cases c1 and c2,
$p_0<0$ and the particle falls into the potential well
towards interior spacelike infinity. A special case of these
traversable wormholes is the symmetrical wormhole \cite{3}
$p_2 = 0$ with $x=z$, $y=h^{-1}x$ (belonging to the case
c2). Then the interior metric has the same form
\be
\rd s^2_- = (\rd t - \omega_- \, \rd \varphi')^2 -
\alpha^2_- \, (r-r_-)^2 \, \rd \varphi'^2 - \rd r^2
\ee
as the exterior metric (3.4), with $\alpha_- = \alpha_+$
arbitrary, $\omega_- = - \omega_+$, $r_- = - r_+$, so that
the full space-time admits the discrete isometry $t
\rightarrow t$, $\varphi \rightarrow -\varphi'$, $r
\rightarrow -r$, which exchanges the two axes at spacelike
infinity.

The Kasner exponents $p_0$ and $p_1$ are complex conjugate
in the domains a and d. The matrix $\lambda^{-1}$ is
obtained by inverting eq.\  (2.22), with $\rho = (1-p_2)/2$,
$\sigma^2 = (p_2-1)(3p_2+1)/4$, leading to the geodesic
equation
\be
\left( \frac{\rd r}{\rd \tau}\right)^2 + b \, |r-r_-|^{-
2p_2} + c \, \sin \, (2 \sigma \, \ln |r-r_-| + d) \, |r-r_-
|^{p_2-1} + \eta = 0 \, ,
\ee
with $b \geq 0$. Incoming test particles are always
reflected back to exterior infinity by the effective
potential in (5.9), generically by the potential barrier
$b|r-r_-|^{-2p_2}$ for $p_2<-1/3$, or by the oscillating
term in $|r-r_-|^{p_2-1}$ for $p_2>1$.

Up to now, we have assumed that the interior metric has the
generic form (3.6), excluding the exceptional metrics
(2.26). For an exceptional metric, the matrix $(1/r_-)A_-$
in (3.10) should be replaced by $(1/2)A_-$, where $A_-$ has
the form
\be
A_- = a \, \left(
\begin{array}{cc}
A_2 & 0 \\
0 & 1
\end{array}
\right) \, ,
\ee
with
\be
{\rm Tr}A_2 = {\rm Tr}A_2^2 = -1
\ee
from eq.\  (2.24), and
\be
a = \tau_z - \mu - \tau_\varphi \, .
\ee
The constraints (5.11) lead to the two relations
\ba
& & \frac{1}{r_+} = \frac{1}{2} \, (\mu + \tau_\varphi +
\tau_z) \, , \nonumber \\
& & \beta^2 = \frac{(\mu-\tau_z)^2 + 3 \, \tau_\varphi^2}{3
\, \mu^2 + (\tau_\varphi - \tau_z)^2} \, ,
\ea
which may also be obtained from eq.\  (3.11) by taking the
limit $r_- \rightarrow \infty$; the conditions $r_+<0$,
$\beta <1$ are satisfied if
\be
\mu < \tau_\varphi < - \mu - \tau_z \, .
\ee
If $a \neq 0$ ($\tau_z \neq \mu + \tau_\varphi$) we obtain,
using (2.27) and (5.10), the interior metric
\be
\rd s_-^2 = {\rm e}^{-ar/2} \, (C_2 \, {\rm
e}^{B(\sqrt{3}/2)ar})_{ab} \, \rd x^a \, \rd x^b - {\rm
e}^{ar} \, \rd z^2 - \rd r^2 \, ,
\ee
where $C_2$ and $B$ are $2 \times 2$ matrices, with $B^2 = -
1$. Again, we can show that incoming test particles are
reflected back to exterior infinity by an exponentially
rising potential barrier if $a>0$, an oscillating potential
barrier if $a<0$. In the special case $a=0$, corresponding
to
\be
\tau_\varphi = \pm \beta \, \mu \, , \,\,\, \tau_z = (1 \pm
\beta) \, \mu \, ,
\ee
with $\mu<0$, $A$ is of rank $1$ and the interior metric
\cite{3}
\ba
\rd s_-^2 & = & (\gamma r + 1) \, (\rd t - \omega \, \rd
\varphi)^2 - 2 \, \varepsilon'' \, \alpha \, r_+ \gamma r \,
(\rd t - \omega \, \rd \varphi) \, \rd \varphi \nonumber \\
& & \mbox{} + \alpha^2 \, r_+^2 \, (\gamma r - 1) \, \rd
\varphi^2 - \rd z^2 - \rd r^2 \, ,
\ea
(with $\gamma = \pm 2 \beta \mu /(1 \pm \beta)$,
$\varepsilon'' = \pm {\rm sign} \Omega$) is flat. In the
case of the sign $+$ in (5.16), $\gamma<0$ and incoming test
particles are reflected by a linearly rising potential,
while for the sign $-$ the particles fall towards interior
infinity with constant proper acceleration \cite{3}.
Finally, if $\beta=0$ in (5.16), $A$ is of rank $0$ and the
interior metric reduces to the Minkowski cylinder (2.34).
\section{Discussion}

We have determined and analyzed the various possible regular
space-time metrics inside an infinite rotating hollow
cylinder, the metric outside the cylinder being chosen of
the spinning cosmic string type. While the analogous
spherically symmetric problem admits a unique solution, a
traversable wormhole made of two Schwarzschild space-times
with a spherical matter shell as common boundary \cite{20},
the variety of solutions we have obtained here is a striking
 illustration of the non-existence of a Birkhoff theorem in
the case of cylindrical symmetry. These solutions include a
space-time with topologically Euclidean spatial sections,
eq.\  (4.4), a black hole solution, eq.\  (4.17), a quasi-regular
solution, eq.\  (4.27) with $a<0$, as well as the
wormhole solutions of section 5. The special solution (4.35)
might approximately describe spontaneous compactification of
the cylinder to a torus.

Although the exterior metric is the same for all these
solutions, an observer at exterior spatial infinity can gain
some information on the values of the cylinder parameters by
aiming beams of test particles towards the cylinder. In the
case of the black-hole solution, as well as for the
traversable wormhole solutions c1 and c2 of section 5 and
the exceptional solution (5.17) with the sign $-$, all the
incoming test particles disappear inside the cylinder. Some
photons may be absorbed by the cylinder in the case of the
 wormhole solutions b and c3, while all incoming test
particles are reflected back to the outside in the other
cases. If we insist on the positivity of the cylinder energy
density $\mu$, then only the topologically Euclidean
solution (4.4), the quasi-regular solution (4.27) and the
exceptional wormhole solution (5.15) remain. In these cases
incoming test particles are always reflected back, so that
total or partial absorption of the incident beam would
signal that the matter making up the cylinder is exotic.

After this work was completed, we came across an article by
Khorrami and Mansouri \cite{8}. The hollow cylinder considered
by these authors is non-rotating ($\beta = 0$), but may collapse
or expand. Assuming \footnote{The purported demonstration of eq.
(31) is flawed.} the exterior metric to be the static cosmic
string metric, and choosing the interior metric to be the
topologically Euclidean metric (4.4), the authors of \cite{8}
recover our relation (4.1) $\tau_z = \mu + \tau_\varphi$, and
find that the static solution corresponds to $\tau_\varphi = 0$
(which results from our eq. (4.7) with $\beta = 0$).

\vspace{5 mm}
\large \noindent Acknowledgement
\vspace{5 mm}

\normalsize \noindent Part of this work was carried out
while G.\ C.\ was at the Institut Non Lin\'{e}aire de Nice
(INLN). I.\ Z.\ wishes to thank the INLN and Professor
F.~Rocca (Universit\'{e} de Nice) for the kind hospitality
afforded during part of this work.

\newpage
\begin{figure}
\vspace{10 cm}
\caption{Penrose diagram for the black-hole space-time
generated by two mirror-symmetric matter cylinders (heavy
curved lines). The exterior regions are truncated conical
space-times, while the interior region is a truncated Rindler
cylinder.}
\end{figure}
\end{document}